%% file: arxiv.tex
\documentclass[11pt]{article}
\usepackage{arxiv}
\usepackage[utf8]{inputenc}
\usepackage{amsmath,amssymb}
\usepackage{graphicx}
\usepackage{booktabs}
\usepackage{multirow}
\usepackage{xcolor}
\usepackage{hyperref}
\usepackage{cite}
\usepackage{tabularx}
\usepackage{siunitx}
\usepackage{subcaption}
\usepackage[capitalize]{cleveref}
\usepackage{tikz}
\usepackage{svg}
\usetikzlibrary{positioning,arrows.meta,fit,backgrounds,calc,decorations.pathreplacing}
\usepackage{algorithm}
\usepackage{algorithmic}
\usepackage{booktabs}

\hypersetup{colorlinks=true, linkcolor=blue, citecolor=blue, urlcolor=blue}

\title{Dynamic Ultrasound Beamforming Using Left-to-Right Arithmetic Adders on FPGA}
\usepackage{authblk}

\author[1]{Muhammad Usman}
\author[2]{Shujaat Khan}
\author[1]{Dorit Merhof}

\affil[1]{Faculty of Informatics and Data Science, University of Regensburg, 93053 Regensburg, Germany \protect\\ \texttt{\{muhammad.usman,dorit.merhof\}@ur.de}}
\affil[2]{Department of Computer Engineering, College of Computing and Mathematics, KFUPM, \protect \\Dhahran, 31261, Saudi Arabia \protect\\ \texttt{shujaat.khan@kfupm.edu.sa}}

\begin{document}
% \maketitle
\begingroup
\renewcommand\thefootnote{}
\footnotetext{Accepted for publication at IEEE 39th International System-on-Chip Conference (SOCC)}
\endgroup

\maketitle
\input{0_Abstract}


\input{1_Introduction}

\input{3_Prroposed_Architecture_Arxiv}

\input{2_Evaluation_Setup}

\input{4_Results}

\input{5_Conclusion}

\input{6_Ack}

\input{Ref}
\end{document}

%% file: 0_Abstract.tex
\begin{abstract}
Adder trees are the computational backbone of delay-and-sum (DAS) ultrasound beamforming, where their implementation directly determines the energy, throughput, and area of a real-time imaging pipeline. Conventional parallel adder trees perform full-precision combinational reduction on every sample, leading to wide critical paths, high LUT consumption, and timing failures on small FPGA devices. This paper presents an alternative adder tree architecture based on \emph{left-to-right (LR)} or \emph{most significant digit first (MSDF) arithmetic}. We implement the proposed and conventional adder trees on a Xilinx Zynq XC7Z010 FPGA and evaluate them for DAS beamforming of a 64-channel ultrasound dataset. The proposed design uses 2.5$\times$ fewer LUTs than the smallest conventional tree, successfully meets the timing constraint, and consumes 23\% less dynamic power than the most efficient conventional baseline.
A key advantage of the proposed MSDF adder tree is that it can generate high-quality beamformed images without waiting for full-precision completion. This naturally enables dynamic precision at runtime with negligible control overhead, since precision selection is achieved simply by stopping the computation clock after the desired number of cycles. Such quality--energy scalability is fundamentally unavailable in conventional fixed-cycle adder trees.
Iso-area replication enables up to 15 parallel instances on the XC7Z010, achieving 67 FPS, which is 80\% higher throughput than the best conventional design.
\end{abstract}

%% file: 1_Introduction.tex
%===================================================================
\section{Introduction}\label{sec:intro}

Ultrasound imaging is widely used in clinical practice because it provides real-time visualization, portability, and safe operation without ionizing radiation~\cite{szabo2013diagnostic}. In a conventional B-mode system, a transducer array acquires raw radio-frequency (RF) channel data, which must then be beamformed to reconstruct the final image. Delay-and-sum (DAS) remains the most widely used beamforming method because of its simplicity, effectiveness, and suitability for real-time imaging. The full DAS receive pipeline encompasses delay calculation, sample interpolation, and coherent summation across all active channels. While adaptive beamformers and deep-learning-based reconstruction methods can improve image contrast and spatial resolution beyond conventional DAS~\cite{luijten2023ultrasound,yan2023fast,malamal2024fpga}, their substantially higher computational complexity renders real-time deployment on compact embedded platforms impractical. DAS therefore remains the de facto standard beamforming kernel for portable and FPGA-based ultrasound systems, and its efficient hardware realization is a continuing research priority.

\begin{figure}
    \centering
    % \fbox{
    \includegraphics[width=1.0\columnwidth]{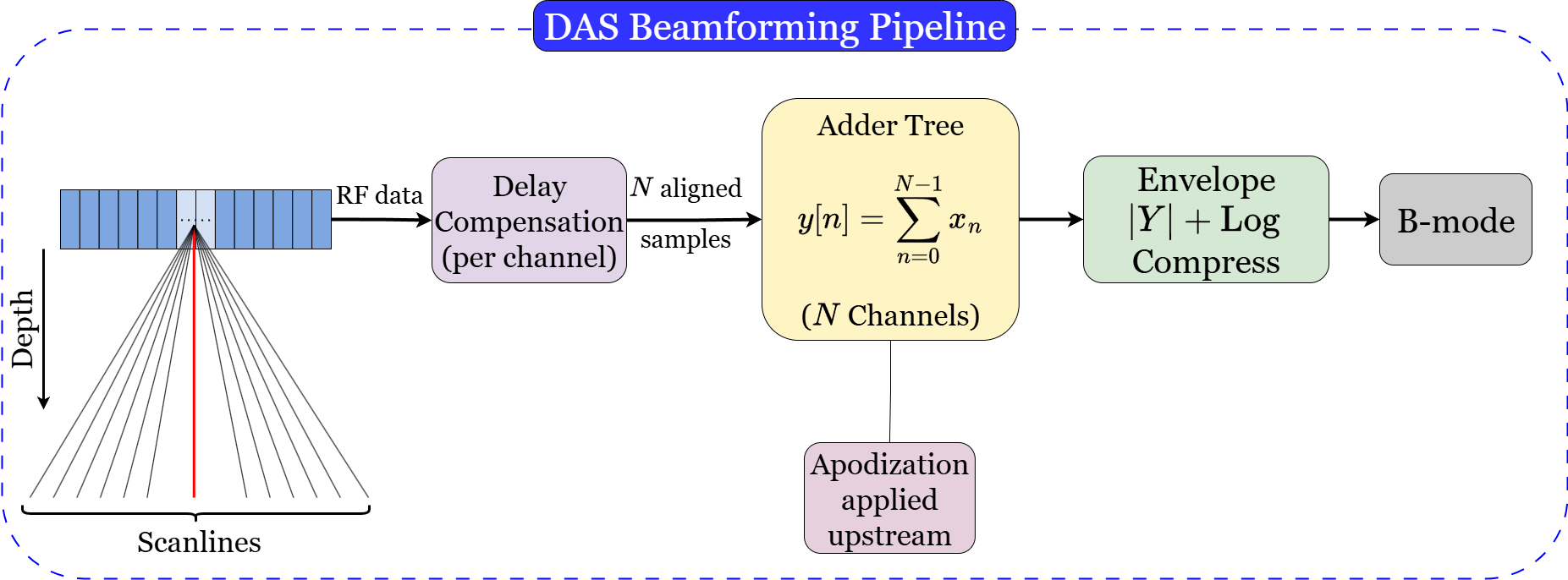}
    % }
    \caption{DAS beamforming pipeline, where channel samples
are delay-aligned and summed to form beamformed pixels
across scanlines}
    \label{fig:beamformer}
\end{figure}

As illustrated in Fig.~\ref{fig:beamformer}, DAS beamforming forms each output pixel by first applying the appropriate focusing delays to the received channel samples and then summing the aligned values across the receive aperture. For an $N$-element array, the beamformed output at pixel $m$ is given by:
\begin{equation}
y(m) = \sum_{n=0}^{N-1} x_n\!\left(m-\Delta_n(m)\right),
\end{equation}
where $x_n(\cdot)$ denotes the RF signal acquired at channel $n$ and $\Delta_n(m)$ is the corresponding focusing delay. Since this accumulation must be repeated for every pixel, channel summation becomes the dominant arithmetic kernel in real-time ultrasound beamforming and a major source of performance and energy cost, particularly on compact FPGA-based systems~\cite{wang2024handheld}. For a 64-channel system producing over 240{,}000 pixel summations per frame, sustaining clinical frame rates of 30--60\,FPS requires the adder tree to perform hundreds of millions of signed additions per second, placing tight demands on both area and power.

FPGA-based beamformers are attractive because they provide deterministic latency, abundant parallelism, and design reconfigurability~\cite{kou2023high,zhang2025ultrafast}. In conventional implementations, the delayed channel samples are reduced by a parallel adder tree within a single clock cycle~\cite{kidav2021fpga}. While this maximizes throughput per cycle, it also creates several challenges on resource-constrained FPGA devices: (i) the adder tree may occupy a substantial number of LUTs, (ii) the long combinational paths can reduce the maximum operating frequency and complicate timing closure~\cite{meacci2024hdl}, and (iii) the reduction is performed at fixed full precision even though downstream stages, such as envelope detection, may not require all least-significant bits.

To address these limitations, we propose to use left-to-right (LR) arithmetic~\cite{ercegovac_book}, for channel-sample accumulation in the DAS beamforming pipeline. In LR arithmetic, operands are consumed and results are produced from the most-significant digit (MSD), so the most informative part of the output becomes available first \cite{usman2023low}. This MSD-first property enables early termination when full precision is unnecessary. The resulting left-to-right adder (LRA) tree replaces wide combinational accumulation with a compact registered digit-serial structure that is better suited to area- and timing-constrained FPGA implementations.

This paper makes the following contributions:
\begin{enumerate}
    \item We propose an MSD-first left-to-right adder-tree architecture for DAS beamforming, realized as a digit-serial reduction structure with an internal registered pipeline.
    
    \item We conduct a comprehensive FPGA evaluation of exact, approximate, and proposed adder-tree architectures using switching-activity interchange format (SAIF)-based post-synthesis power analysis on a Xilinx Zynq XC7Z010 with a 64-channel ultrasound beamforming workload.
    
    \item We show that the proposed architecture supports run-time dynamic precision through variable-$K$ computation, and we characterize the resulting PSNR, SSIM, and B-mode image quality versus energy trade-off.
    
    \item We show through implementation and iso-area scaling analysis that the proposed design provides a more favorable area-throughput-timing trade-off than conventional adder trees on a resource-constrained FPGA platform.
\end{enumerate}

%% file: 3_Prroposed_Architecture_Arxiv.tex
%===================================================================
\section{Proposed Architecture}\label{sec:arch}
%===================================================================
This section describes the baseline and proposed adder-tree organizations considered in this work. We first summarize the conventional reduction architecture and then present the proposed LRA64 design, including its cell structure, tree organization, output reconstruction, and support for run-time precision control.

\subsection{LRA64 Adder Tree}
\subsubsection{LRA Cell}
The left-to-right adder (LRA) cell in Fig.~\ref{fig:oa_cell}(a) performs digit-serial addition in a most-significant-digit-first fashion using redundant signed-digit operands. Since the most-significant digit can be processed as soon as it becomes available, the cell naturally supports pipelined MSDF accumulation.

Its operation follows a two-step addition scheme in which local logic and pipeline registers are used to suppress long carry propagation and generate intermediate results incrementally. As a result, the cell produces valid output digits with an initial delay of $\delta = 2$ cycles while preserving a very small combinational depth. The implementation uses two full adders and five pipeline registers. Further derivation of the underlying MSDF-addition principle is given in~\cite{ercegovac_book}.

\subsubsection{LRA Tree and Accumulation}
\begin{figure}[!t]
\centering
% \fbox{
\includegraphics[viewport=20 20 720 350,scale=0.45]{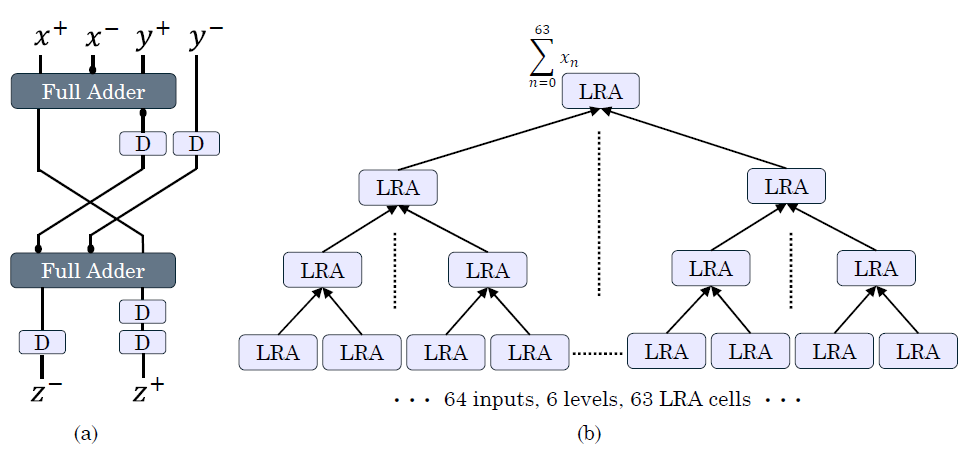}
% }
\caption{Proposed LRA64 architecture. (a) Left-to-right adder (LRA) cell composed of two full adders and five pipeline registers, forming a two-stage SD-2 digit-serial pipeline with initial delay $\delta = 2$. (b) LRA64 adder tree formed by 63 LRA cells organized as a 6-level balanced binary tree for MSD-first accumulation.}
\label{fig:oa_cell}
\end{figure}

As shown in Fig.~\ref{fig:oa_cell}(b), the complete LRA64 reduction network is constructed by arranging the LRA cells in a balanced binary hierarchy. This organization preserves a registered boundary at every level, thereby avoiding the long combinational paths of conventional adder trees. The redundant-digit outputs generated by the tree are finally converted back to a conventional binary sum by a shift-and-add accumulator.

\begin{figure*}[!ht]
\centering
% \fbox{
    \includegraphics[viewport=30 200 690 500,scale=0.65]{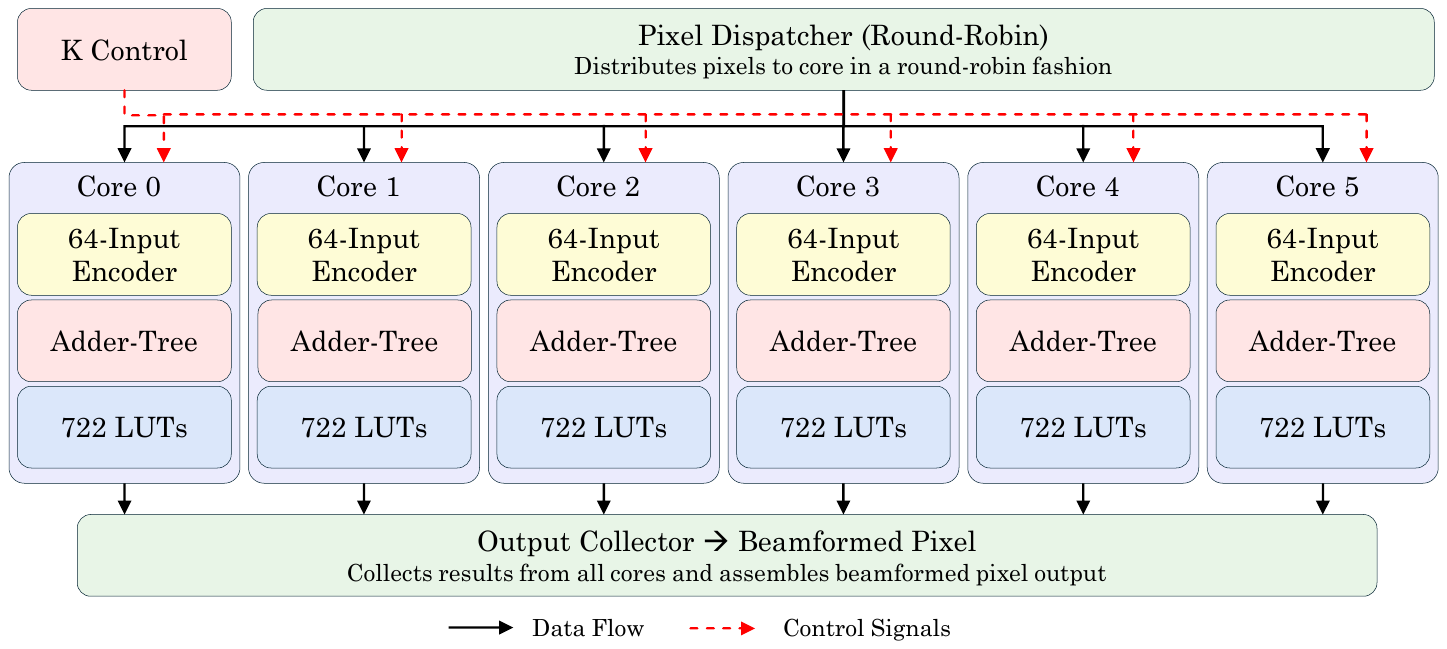}
    % }
\caption{System-level FPGA organization of the proposed LRA64 design. Parallel cores process independent pixel summations under shared precision control. The six illustrated cores use only 25\% of the XC7Z010 LUT budget, while up to 15 instances can be accommodated on the device.}
\label{fig:fpga_system}
\end{figure*}

\subsubsection{MSD-First Encoding and Output Reconstruction}
Before reduction, each signed 16-bit input is converted into an MSD-first digit stream for the LRA tree. The encoder emits the sign contribution first and then serializes the remaining magnitude bits across separate positive and negative channels. After the selected $K$ bit planes have been processed, zeros are injected to flush the registered tree. This encoding step is what enables variable-$K$ operation and keeps the internal computation in a narrow digit-serial form.
The redundant-digit output stream produced by the LRA tree is converted to a conventional binary result by a shift-and-add accumulator:
\begin{equation}
    S_t = 2S_{t-1} + z_p(t) - z_n(t),
    \label{eq:accum}
\end{equation}
where $z_p(t)$ and $z_n(t)$ are the positive and negative output bits generated at cycle $t$. Thus, each cycle shifts the current partial result by one position and incorporates the decoded signed digit. After the selected computation window has completed, the accumulator contains the final binary sum. When all input bit planes are processed, the result matches that of the corresponding full-precision conventional reduction.

\subsection{Dynamic Precision via Early Termination}
The MSD-first operation of LRA64 enables dynamic precision control~\cite{ibrahim2023dslot}. Since the most-significant digit of the result is formed first, the computation may be terminated after processing only the first $K$ bit planes, with $K < 16$. This reduces both cycle count and switching activity, and enables run-time, application-aware precision control on a per-frame or per-mode basis. Such flexibility is not available in conventional single-cycle adder trees.

\subsection{Iso-Area Parallelism}

Owing to its compact implementation, LRA64 can be replicated more aggressively on the XC7Z010 than any conventional adder tree. Under an iso-area scaling analysis, identical instances are instantiated until the FPGA resource budget is exhausted. LRA64 (\num{722} LUTs, \num{2314} FFs) is limited by FF usage and supports up to 15 parallel instances, whereas all conventional baselines are limited by LUT capacity.

This directly improves throughput: the projected performance of LRA64 reaches 66.9 FPS, compared with 38.0 FPS for the best exact conventional baseline (RCA) and 11.0 FPS for Kogge--Stone. The corresponding effective input rate is 1047.7 MSa/s for LRA64, versus 595.3 MSa/s for RCA, yielding a 80\% advantage. The resulting multi-instance organization is illustrated in Fig.~\ref{fig:fpga_system}.

%% file: 2_Evaluation_Setup.tex
\section{Evaluation Setup}
This section discusses about the dataset used in this study, the evaluation framework, including the beamforming kernel, baseline architectures, the flow of FPGA synthesis its power-analysis and the corresponding metrics used to assess image quality and dynamic precision. 

\subsection{Dataset}
The RF dataset used in this study was acquired using an E-CuBE R12 ultrasound system equipped with a linear probe operating at 8.5\,MHz. Data collection was performed on an ATS-549 multipurpose tissue-mimicking phantom, which provides standardized targets for evaluating image quality in terms of resolution, contrast, and spatial accuracy~\cite{khan2020adaptive}.

\subsection{Beamforming Kernel and Workload}
\begin{figure}[t]
    \centering
    \begin{center}
    % \fbox{
    \includegraphics[width=\columnwidth]{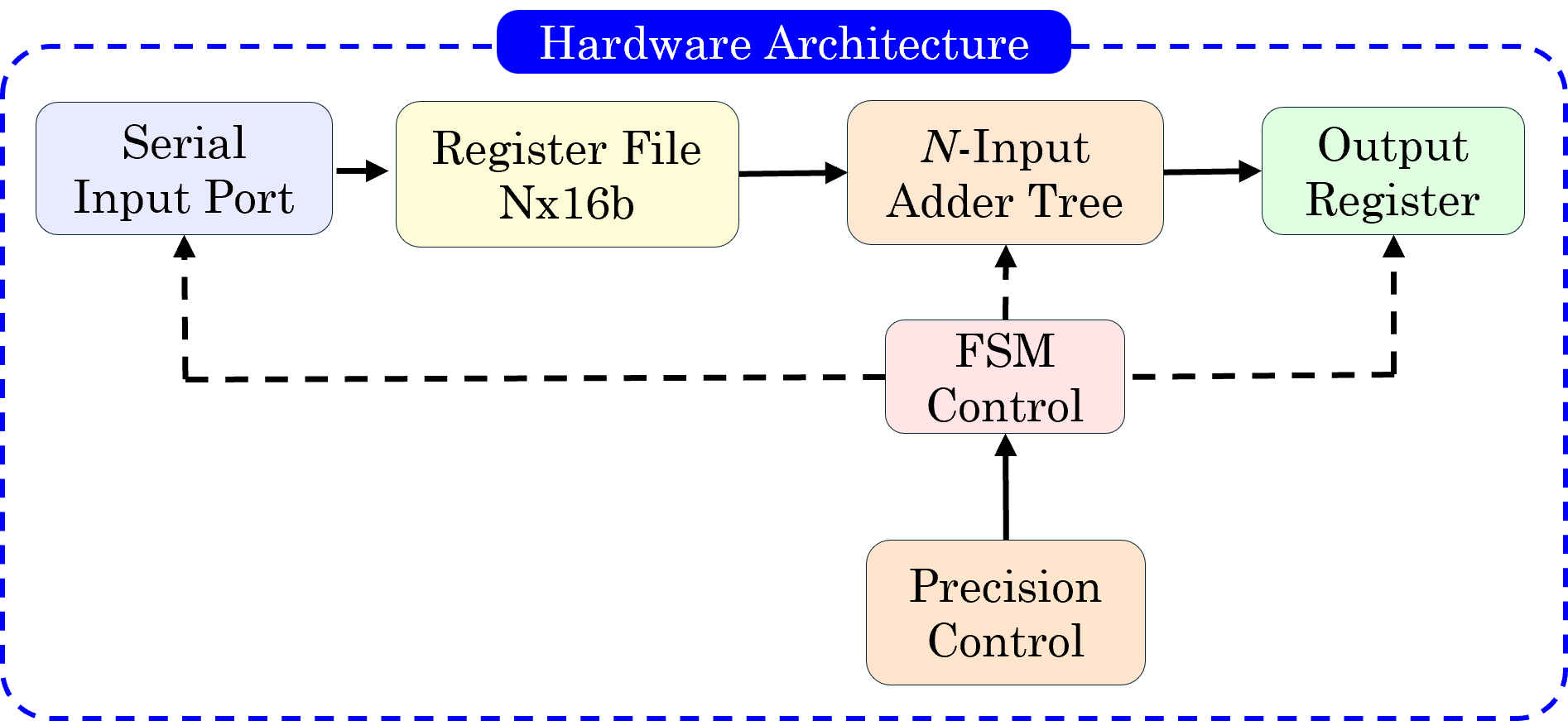}
    % }
    \end{center}
    \caption{Proposed FPGA architecture of the $N$-channel MSD-first adder tree with FSM-based control and run-time precision control.}
    \label{fig:system_overview}
\end{figure}

This work focuses on the channel-accumulation kernel of delay-and-sum (DAS) beamforming. In the considered setup, each beamformed pixel is obtained by summing the delay-compensated samples across all input channels:
\begin{equation}
    S[p] = \sum_{c=0}^{63} x_c[p],
    \label{eq:das}
\end{equation}
where $x_c[p]$ denotes the aligned sample from channel $c$ at pixel $p$. We evaluate this kernel using a 64-channel complex-valued ultrasound RF dataset with \num{122304} pixels. Since the RF data is complex-valued, the real and imaginary components are processed separately, resulting in \num{244608} real-valued channel-summation operations per frame. Figure~\ref{fig:system_overview} provides a high-level view of the corresponding FPGA architecture used to implement this reduction with run-time precision control.

\subsection{Compared Architectures}
The comparison includes seven exact adder-tree architectures, namely Ripple-Carry Adder (RCA), Carry-Lookahead Adder (CLA), Carry-Skip Adder (CSKA), Carry-Select Adder (CSLA)~\cite{jan2003digital}, Brent--Kung~\cite{brent1982regular}, Kogge--Stone~\cite{penchalaiah2018design}, and Sklansky~\cite{sklansky2009evaluation}, together with EvoApprox~\cite{mrazek2017evoapprox8b} as an approximate baseline. Collectively, these designs span a broad range of conventional adder trade-offs. All baselines implement the same LUT-based reduction of signed 16-bit channel samples.

The proposed LRA64 differs from these fixed-precision single-cycle architectures by employing an MSD-first digit-serial adder tree in a balanced binary structure. This organization enables run-time precision control through a configurable parameter $K \in \{1,\dots,16\}$, allowing lower-order bit planes to be omitted without modifying the datapath.

\subsection{Synthesis and Power Measurement}
All designs target the Xilinx Zynq XC7Z010 (\num{17600} LUTs, \num{35200} FFs) in Vivado 2023.2 with a 100\,MHz timing constraint. A common wrapper is used for all architectures to ensure a fair comparison, so that the adder-tree implementation is the only varying component.

\begin{figure}[b] 
\begin{center} 
\centering 
\includegraphics[viewport=70 260 460 500,scale=0.5]{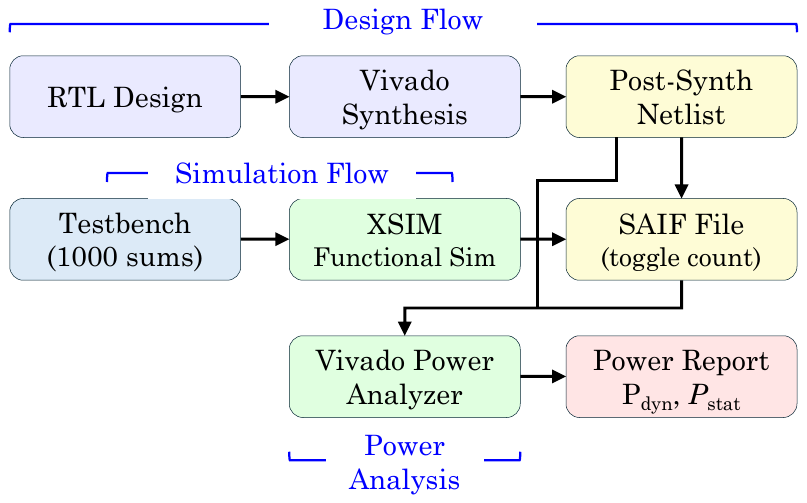} 
\end{center} 
\caption{SAIF-based post-synthesis power analysis flow. RTL is synthesised in Vivado, then a testbench drives the post-synthesis netlist in XSIM to produce a SAIF file with per-net toggle counts. Vivado's power analyzer combines the SAIF data with the netlist to compute accurate dynamic power at high confidence.} 
\label{fig:saif_flow} 
\end{figure}

Power is estimated using SAIF-based post-synthesis simulation. As summarized in Fig.~\ref{fig:saif_flow}, Vivado first generates the post-synthesis netlist, after which a testbench drives representative beamforming summations in XSIM to capture per-net switching activity in SAIF format. This activity is then back-annotated into the power analyzer to report dynamic power. 

% To compare designs fairly despite different cycle counts and achievable clock frequencies, per-image energy is normalized to a common 100\,MHz reference. This metric captures the intrinsic power-latency trade-off of each architecture rather than raw operating frequency alone.

\subsection{Quality Metrics}
Image fidelity is evaluated relative to a 64-bit floating-point reference using normalized mean squared error (NMSE), peak signal-to-noise ratio (PSNR), and structural similarity index (SSIM)~\cite{wang2004image}. 
All exact adder-tree baselines and LRA64 at full precision ($K{=}16$) produce identical 16-bit-quantized outputs. For the dataset used in this study, the approximate baselines likewise introduce no observable difference in the final quantized images, indicating that their deviations are confined to low-order intermediate bits.

\subsection{Dynamic Precision Sweep}
To evaluate the run-time precision capability of LRA64, we sweep $K$ from 1 to 16 and measure the resulting energy and image quality. For a given setting, the compute phase requires $K{+}12$ cycles, while the total latency per result is $K{+}76$ cycles including the 64-cycle load phase.

%% file: 4_Results.tex
%===================================================================
\section{Results and Discussion}\label{sec:results}
%===================================================================

\subsection{Overall FPGA Comparison}
The evaluated adder-tree designs exhibit clear trade-offs among area, timing, energy, and throughput, with the proposed LRA64 providing the most favorable overall balance on the target FPGA, as summarized in \cref{tab:addertree_comparison}. The comparison includes exact, approximate, and proposed MSDF-based designs, and reports resource usage, achievable clock frequency, power, energy, and throughput under both single-instance and iso-area multi-instance deployment.

\begin{figure}[!t]
\centering
\includegraphics[width=\columnwidth]{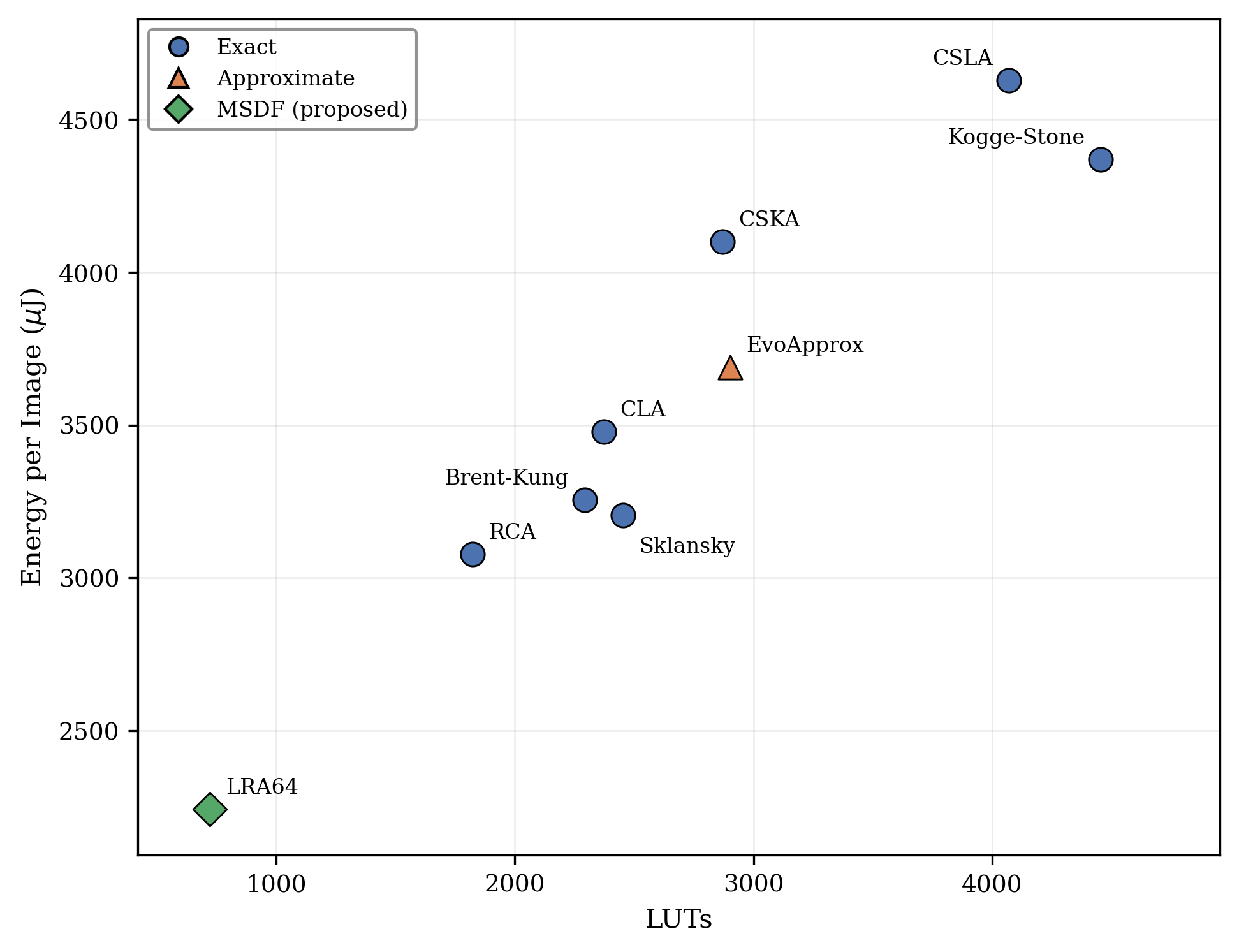}
\caption{Area--energy scatter plot for all nine designs. LRA64 (green diamond) occupies the low-area corner at 722\,LUTs while delivering competitive per-image energy (\SI{2242.29}{\micro\joule}), making it the Pareto-optimal choice for resource-constrained FPGA targets.}
\label{fig:scatter}
\end{figure}

LRA64 achieves the smallest LUT footprint, the highest operating frequency, and the lowest power and energy per pixel among all evaluated architectures. It consumes 10\,mW dynamic power (23\% below RCA, 41\% below CSLA) since digit-serial processing switches only one bit plane per cycle.
The per-image energy of \SI{2242.29}{\micro\joule} is highly competitive, achieving 27\% lower energy than RCA, 14\% lower than Kogge--Stone, and 19\% lower than CSLA. As shown in \cref{fig:scatter}, the area--energy tradeoff places LRA64 at the Pareto-optimal low-area corner.

\renewcommand{\arraystretch}{1.5}
\begin{table*}[t]
\centering
\caption{Comparison of exact, approximate, and proposed adder-tree designs on Xilinx Zynq XC7Z010 (best results indicated in bold).}
\label{tab:addertree_comparison}
\scriptsize
\setlength{\tabcolsep}{3.5pt}
\renewcommand{\arraystretch}{1.08}
\resizebox{\textwidth}{!}{%
\begin{tabular}{llccccccccc}
\toprule
\textbf{Design} & \textbf{Type} & \textbf{LUTs}$\downarrow$ & \textbf{FFs}$\downarrow$ & \textbf{$F_{\max}$ (MHz)$\uparrow$} & \textbf{Dyn. (mW)}$\downarrow$ & \textbf{Energy (pJ/pixel)$\downarrow$} & \textbf{Energy ($\mu$J/img)$\downarrow$} & \textbf{FPS (1 inst.)$\uparrow$} & \textbf{Max Inst.$\uparrow$} & \textbf{FPS$_{\max}\uparrow$} \\
\midrule
RCA              & Exact       & 1823         & 1052 & 69.3           & 13          & 12577.2         & 3076.49         & 4.23          & 9              & 38.03 \\
CLA              & Exact       & 2374         & 1052 & 66.0           & 14         & 14222.0         & 3478.80         & 4.02          & 7              & 28.17 \\
CSKA             & Exact       & 2871         & 1056          & 56.0           & 14          & 16763.0         & 4100.36         & 3.41          & 6              & 20.49 \\
CSLA             & Exact       & 4068         & \textbf{1049} & 60.2           & 17          & 18923.3         & 4628.80         & 3.67          & 4              & 14.69 \\
Brent--Kung      & Exact       & 2374         & 1052 & 66.0           & 13          & 13206.1         & 3230.32         & 4.02          & 7              & 28.17 \\
Kogge--Stone     & Exact       & 4454         & \textbf{1049} & 60.0           & 16          & 17857.4         & 4368.06         & 3.66          & 3              & 10.99 \\
Sklansky         & Exact       & 2374         & 1052 & 66.0           & 13          & 13206.1         & 3230.32         & 4.02          & 7              & 28.17 \\
EvoApprox        & Approximate & 2903         & 1052 & 62.2           & 14          & 15084.9         & 3689.89         & 3.79          & 6              & 22.76 \\
LRA64 (Proposed) & MSDF        & \textbf{722} & 2314          & \textbf{100.4} & \textbf{10} & \textbf{9166.9} & \textbf{2242.29} & \textbf{4.46} & \textbf{15}    & \textbf{66.90} \\
\bottomrule
\end{tabular}%
}
\end{table*}

\subsection{Dynamic Precision Trade-off and B-Mode Quality}

A key capability of LRA64 is its support for dynamic precision control. Because computation proceeds MSD-first, accumulation can be terminated once sufficient output fidelity has been reached, discarding only the least-significant bits. This is fundamentally different from conventional adder trees, which must complete the full combinational reduction regardless of the required output quality.

\begin{figure}
    \centering
    \includegraphics[width=\columnwidth]{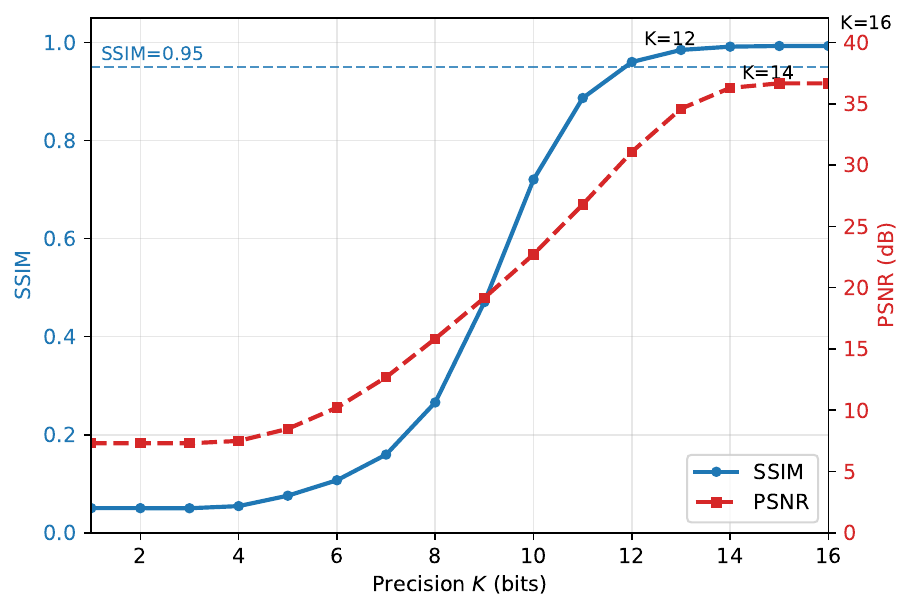}
    \caption{PSNR and SSIM as functions of precision $K$, obtained by applying LRA64-computed sums to a simulated point-scatterer phantom and comparing the reconstructed B-mode images against a float64 DAS reference.}
    \label{fig:quality_precision}
\end{figure}

The quality trend across the precision range is shown in Fig.~\ref{fig:quality_precision}, which reports PSNR and SSIM versus $K$ for a simulated point-scatterer phantom relative to a float64 DAS reference. Image fidelity improves steadily as $K$ increases, with the most pronounced gain appearing between $K{=}8$ and $K{=}12$. Selected operating points are listed in Table~\ref{tab:sweep}. At $K{=}14$, LRA64 achieves PSNR\,=\,36.30\,dB and SSIM\,=\,0.992, while still saving 2.2\% energy relative to $K{=}16$. At $K{=}12$, the design reaches SSIM\,=\,0.960 with 4.3\% energy savings. Below $K{=}10$, however, image quality degrades rapidly, making such settings more suitable for preview-quality or strongly energy-constrained modes.

\begin{table}[t]
\centering
\caption{LRA64 dynamic-precision sweep. Energy savings are relative to full precision ($K{=}16$).}
\label{tab:sweep}
\small
\begin{tabular}{@{}rrrrrr@{}}
\toprule
$K$ & Comp.\ cyc & Total cyc & PSNR (dB) & SSIM & Saving (\%) \\
\midrule
1  & 13 & 77 & 7.30  & 0.050 & 16.3 \\
4  & 16 & 80 & 7.49  & 0.054 & 13.0 \\
8  & 20 & 84 & 15.83 & 0.266 & 8.7 \\
10 & 22 & 86 & 22.72 & 0.721 & 6.5 \\
12 & 24 & 88 & 31.08 & 0.960 & 4.3 \\
14 & 26 & 90 & 36.30 & 0.992 & 2.2 \\
16 & 28 & 92 & 36.67 & 0.993 & 0.0 \\
\bottomrule
\end{tabular}
\end{table}

\begin{figure}[t]
\centering
% \fbox{
% \includegraphics[viewport=70 175 550 490,scale=0.50]{figures/bmode_condensed.pdf}
\includegraphics[width=\columnwidth]{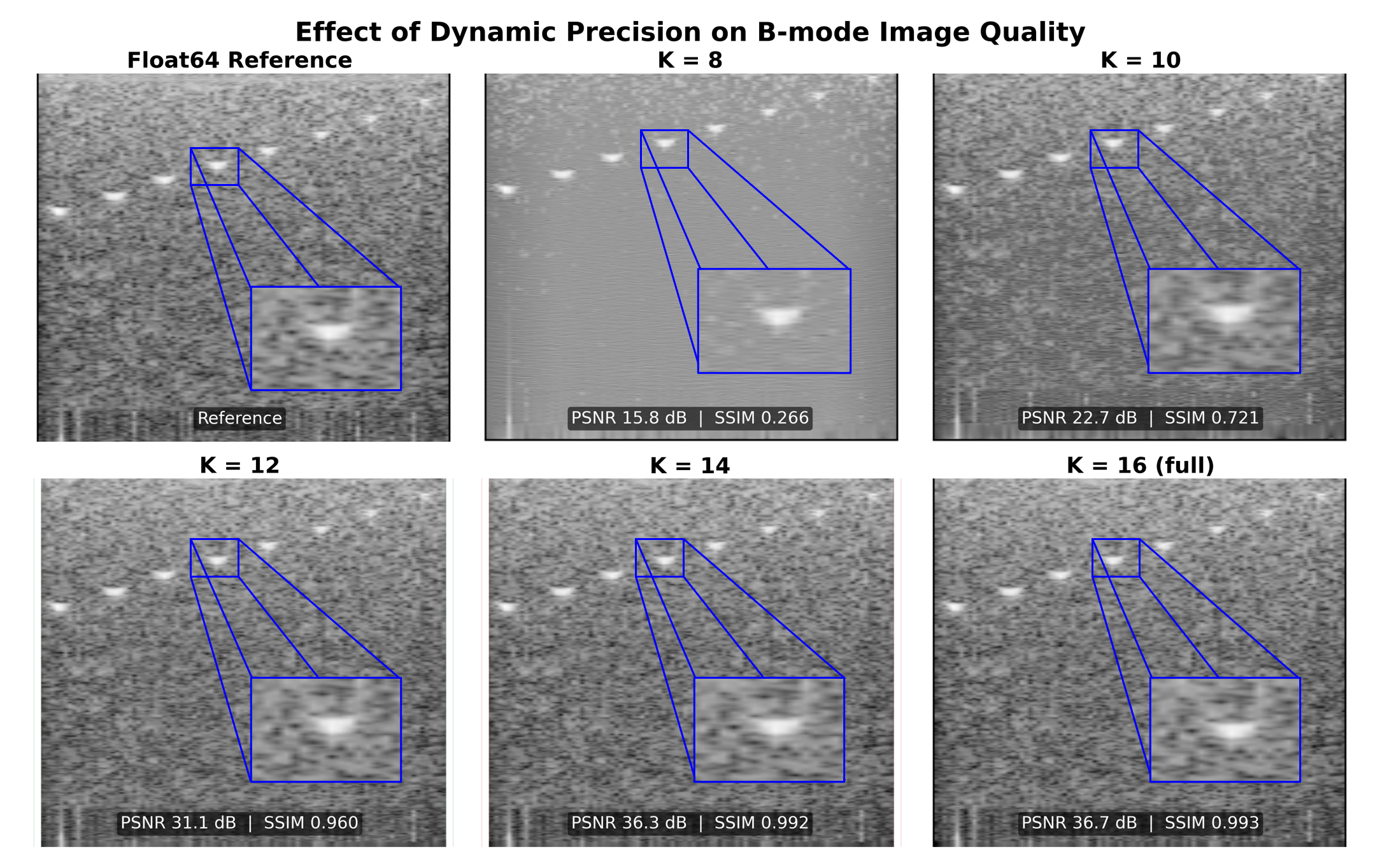}
% }
\caption{B-mode image quality at six precision levels (point-scatterer phantom). Row~1: Float64 reference, $K=8$ (noise-dominated, SSIM\,=\,0.107), $K=10$ (preview quality, SSIM\,=\,0.721). Row~2: $K=12$ (SSIM\,=\,0.960, above 0.95 threshold), $K=14$ (near-reference, SSIM\,=\,0.992), $K=16$ full precision (SSIM\,=\,0.993).}
\label{fig:bmode}
\end{figure}

This trend is also evident in the reconstructed B-mode images shown in \cref{fig:bmode}. At low precision, the image is dominated by noise and lacks coherent structure, whereas scatterer features begin to emerge at intermediate $K$. High structural fidelity is recovered for $K \geq 12$, and the reconstruction at $K{=}14$ is visually almost identical to the full-precision output. These results demonstrate that LRA64 provides a continuous run-time precision knob, enabling the trade-off between image quality and implementation efficiency without hardware reconfiguration.

%% file: 5_Conclusion.tex
%===================================================================
\section{Conclusion}\label{sec:conc}
%===================================================================
This paper presented a comprehensive comparison of adder-tree architectures for the DAS ultrasound beamforming kernel on a Xilinx Zynq XC7Z010. All designs implement the same core operation, namely the accumulation of 64 signed 16-bit channel samples, thereby isolating the effect of the underlying adder-tree architecture on area, timing, energy, and throughput. The results show that the proposed left-to-right adder tree provides the most favorable overall trade-off on this resource-constrained FPGA platform. It achieves the smallest area at 722 LUTs, is the only design that satisfies the 100\,MHz timing target, and maintains per-image energy comparable to the best exact conventional alternatives at full precision. More importantly, unlike conventional fixed-precision trees, LRA64 enables run-time dynamic precision control, allowing a continuous quality-energy trade-off with near-lossless image quality at $K=12$ and further energy reduction at lower precision settings. Under iso-area scaling, this efficiency translates into higher achievable throughput through multi-instance parallelism. These results highlight LR arithmetic as a promising direction for real-time and energy-aware ultrasound beamforming on compact FPGA devices. Future work will extend the approach to volumetric beamforming, include post-implementation power analysis, and investigate adaptive per-pixel precision control.

%% file: 6_Ack.tex
\section*{Acknowledgement}\label{sec:Ack}
This work was supported by the German Research Foundation
(Deutsche Forschungsgemeinschaft, DFG) under project number `573796083'.

%% file: Ref.tex
%===================================================================
% References
%===================================================================
\bibliographystyle{IEEEtran}
\bibliography{reference}